# Gate controlling of quantum interference and direct observation of anti-resonances in single molecule charge transport

Yueqi Li[1], Marius Buerkle[2], Guangfeng Li[3], Ali Rostamian[1], Hui Wang[4], Zixiao Wang[4], David R. Bowler[5,6], Tsuyoshi Miyazaki[6], Limin Xiang[1], Yoshihiro Asai[2]*, Gang Zhou[3]* and Nongjian Tao[1,4]*

Quantum interference can profoundly affect charge transport in single molecules, but experiments can usually measure only the conductance at the Fermi energy. Because, in general, the most pronounced features of the quantum interference are not located at the Fermi energy, it is highly desirable to probe charge transport in a broader energy range. Here, by means of electrochemical gating, we measure the conductance and map the transmission functions of single molecules at and around the Fermi energy, and study signatures associated with constructive and destructive interference. With electrochemical gate control, we tune the quantum interference between the highest occupied molecular orbital and lowest unoccupied molecular orbital, and directly observe anti-resonance, a distinct feature of destructive interference. By tuning the molecule in and out of anti-resonance, we achieve continuous control of the conductance over two orders of magnitude with a subthreshold swing of ~17 mV dec$^{-1}$, features relevant to high-speed and low-power electronics.

As devices decrease to the scale of the electronic phase coherence length, the wave nature of electrons becomes evident[1,2]. Studying single electron interference provides key insights into quantum transport in molecules, and manipulating single molecule conductance at the level of the wavefunction promises device functions[3,4] that are distinctly different from conventional electronics. The interference between the highest occupied molecular orbital (HOMO) and lowest unoccupied molecular orbital (LUMO) is often the most dominant effect on the conductance of a single molecule[5,6]. Based on this, orbital rules have been proposed theoretically[7–9] and examined experimentally[10–14], showing constructive interference in linear conjugated (or *para* oriented) structures and destructive interference in cross-conjugated (or *meta* oriented) structures. However, these studies have focused on demonstrating quantum interference by designing molecules with different structures or modifying the structures via chemical reactions[3,12,15–17]. Here we show that the charge transport determined by the quantum interference in a single molecule can be tuned continuously with an external electrochemical gate without changing the molecular structures (Fig. 1a). This allows us to achieve gate control of single molecule conductance based on quantum interference, rather than carrier densities as in conventional field-effect transistors, to measure the transmission function of the molecule and to study directly the most pronounced quantum interference features, including the destructive interference characteristic anti-resonance, in the transmission function.

We studied a *meta*-oriented diphenyl benzene structure (denoted *Meta*), where substituents occupy positions 1 and 3, and a *para*-oriented diphenyl benzene structure (denoted *Para*), where substituents occupy opposite sites (Fig. 1b). These molecules were connected to the two electrodes and studied using a scanning tunnelling microscope (STM) break junction technique[18] (Fig. 1a). We tuned the quantum interference in the molecules with an electrochemical gate voltage, determined the conductance versus gate voltage (transmission functions) and current–voltage characteristic curves, performed theoretical calculations of the transmission functions, and compared the results with the experiments.

The conductance ($G$) of a single molecule junction at small bias voltages is given by the Landauer formula:

$$G(E_F) = \frac{2e^2}{h} T(E_F) \quad (1)$$

where $e$ is the electron charge, $h$ is Planck's constant and $T(E_F)$ is the electron transmission function at the electrode Fermi energy, $E_F$. Quantum interference affects the transmission function, and thus the conductance of the molecule. For the *Para* and *Meta* molecules, the quantum effect can be understood intuitively by including only the frontier orbitals (HOMO and LUMO) in the zeroth-order Green function, written as[6,7]

$$\frac{C_{H(\beta)} C^*_{H(\alpha)}}{E_F - \varepsilon_H} + \frac{C_{L(\beta)} C^*_{L(\alpha)}}{E_F - \varepsilon_L} \quad (2)$$

where the first and second terms are the zeroth-order Green functions of HOMO and LUMO, respectively, $C_{H,L(\alpha)}$ and $C_{H,L(\beta)}$ are the molecular orbital coefficients of the HOMO (H) and LUMO (L) at the two anchoring atoms, α and β, connected to the two electrodes, and $\varepsilon_{H,L}$ represents the molecular orbital energies of the HOMO (H) and LUMO (L), respectively. The square of the Green function is roughly proportional to the transmission function (Supplementary equation (2)) and the interference is determined by the inter-orbital coupling induced by the electrodes[11]. By considering the phases of the frontier orbitals at the anchoring atoms shown in Fig. 1c, we conclude that the signs of the numerators in equation (2) to be opposite for *Para* and the same for *Meta*. The signs of the denominator on the other hand are always opposite given that the Fermi energy lies within the HOMO–LUMO gap[11]. Hence, the HOMO and LUMO contributions to the conductance add up for *Para*, resulting in constructive interference, and cancel out for *Meta*, leading to destructive interference.

[1]Biodesign Center for Bioelectronics and Biosensors, Arizona State University, Tempe, AZ, USA. [2]Research Center for Computational Design of Advanced Functional Materials (CD-FMat), National Institute of Advanced Industrial Science and Technology (AIST), Tsukuba, Ibaraki, Japan. [3]Laboratory of Advanced Materials, State Key Laboratory of Molecular Engineering of Polymers, Fudan University, Shanghai, China. [4]State Key Laboratory of Analytical Chemistry for Life Science, School of Chemistry and Chemical Engineering, Nanjing University, Nanjing, China. [5]London Centre for Nanotechnology and Department of Physics and Astronomy, University College London, London, UK. [6]International Centre for Materials Nanoarchitectonics (MANA), National Institute for Materials Science (NIMS) Tsukuba, Ibaraki, Japan. *e-mail: yo-asai@aist.go.jp; zhougang@fudan.edu.cn; nongjian.tao@asu.edu

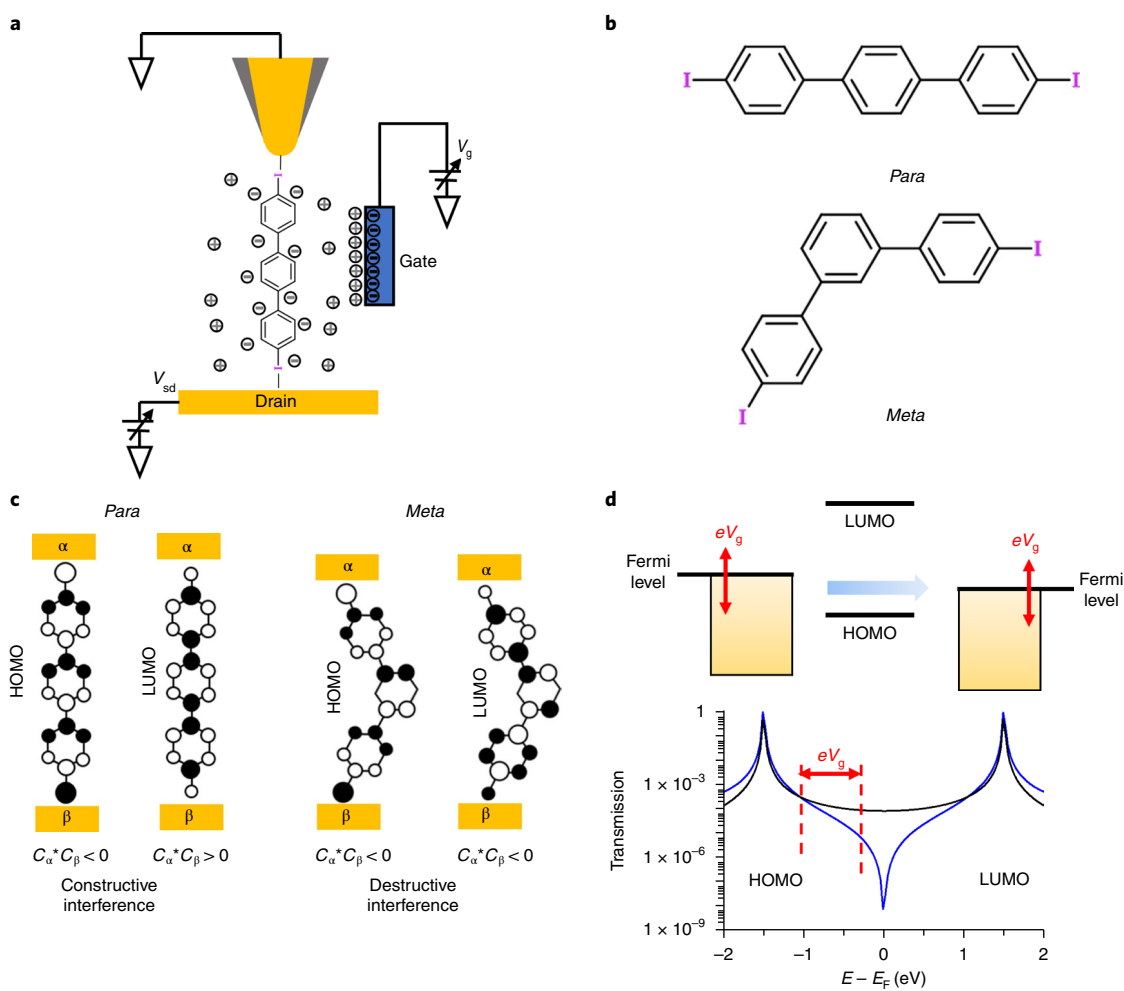

**Fig. 1 | *Para* and *meta* molecules measured by a gate-controlled STM break junction. a**, Schematic view of the gate-controlled measurement of single molecule charge transport. **b**, Chemical structures of the molecules with constructive quantum interference (*Para*) and destructive quantum interference (*Meta*). **c**, Schematic view of the frontier orbitals of *Para* and *Meta*. The size of the circles represents the amplitude of the molecular orbitals. Filled black and white colours represent the different phases of the molecular orbitals. Here, the orbital coefficients $C_{\alpha,\beta}$ are assumed to be real. The asterisks mean complex conjugate. **d**, Top: adjusting the gate voltage shifts the molecular energy levels relative to the electrode's Fermi energy level, which varies the occupancy of the two interfering charge transport channels. Bottom: adjusting the gate voltage changes the relative alignment between molecular energy levels and Fermi level, allowing probing of the energy dependence of transmission. Black and blue curves are representative transmission functions of constructive (black) and destructive (blue) interference situations, respectively.

When a gate voltage $V_g$ is applied to the molecule, the Fermi levels of the STM tip and the substrate shift relative to the molecule's HOMO and LUMO levels (Fig. 1d). This shift in the energy level alignment influences the contribution from the inter-orbital coupling term between the HOMO and LUMO[11], and thus the quantum interference in the molecules. In this way, the energy dependence of the transmission functions can be probed by continuously tuning the gate voltage[19]. For *Meta*, an anti-resonance in the transmission function located inside the HOMO–LUMO gap is predicted. As a result, the transmission function of *Meta* displays a strong energy dependence. In contrast, for *Para*, no anti-resonance is expected, and the transmission function is thus less energy dependent. The intuitive model qualitatively agrees with more sophisticated theories[11,20,21], and also with detailed first-principle calculations, as presented in the calculation part.

We note that electrochemical gating has been applied previously to study charge transport in molecules[3,12,15]. In these previous works, however, chemical reactions took place, which changed the molecular structures and accordingly the molecules' HOMO and LUMO energies and wavefunctions. In contrast, the electrochemical gate in the present work changed only the contribution from the HOMO and LUMO inter-orbital coupling term (which determines the quantum interference)[11,22,23], while the structures and the HOMO and LUMO of the molecules remained unchanged. This allowed us to continuously tune the quantum interference in the single molecules studied here.

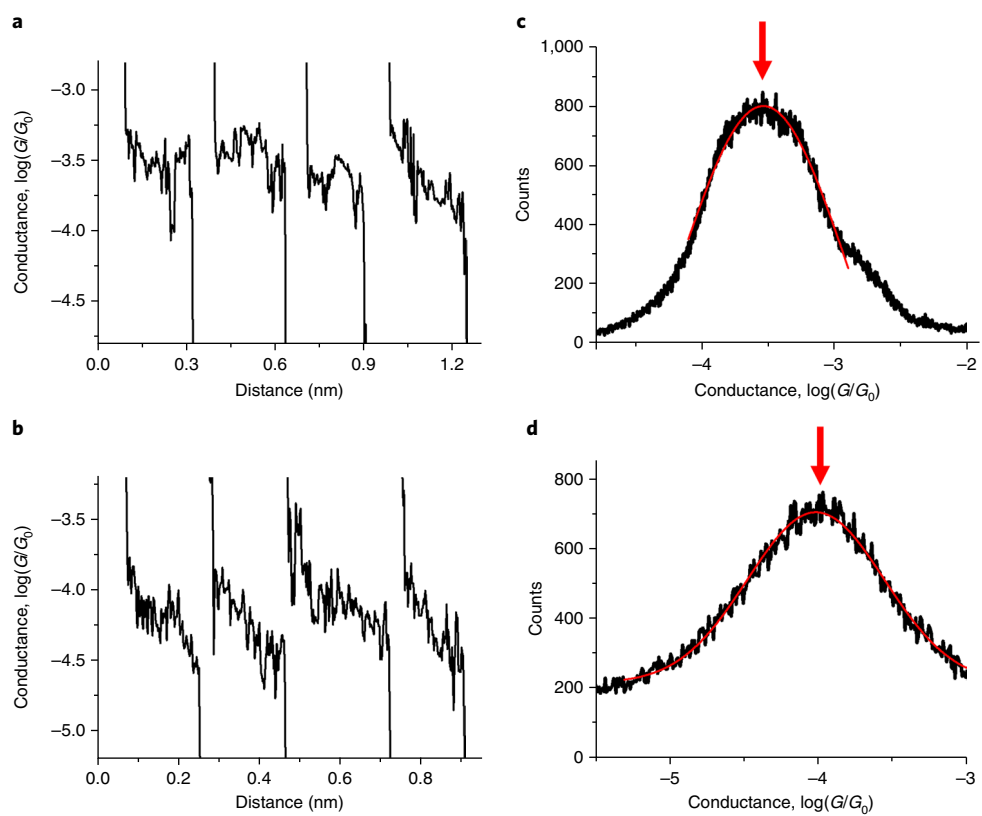

**Fig. 2 | Conductance measurement of *Para* and *Meta* in mesitylene without gate applied. a,b**, Individual conductance versus distance traces recorded from the break junction measurement of *Para* (**a**) and *Meta* (**b**). **c,d**, Conductance histograms of *Para* (**c**) and *Meta* (**d**) constructed from individual curves. Red curves are Gaussian fits of the conductance peaks. Peak positions are marked by red arrows.

Both isomers (*Para* and *Meta*) are terminated with two iodine atoms. Previous works indicated that the iodine terminals could bind with the gold electrodes[24–26] and form molecular junctions (electrode–molecule–electrode)[11,27]. Our X-ray photoelectron spectroscopy, electrochemical and UV–vis data in the present study (Supplementary Fig. 1) also show that the molecules bind with gold surfaces via iodine terminals in a loosely packed form. We performed STM break junction measurements[18,28] using a gold tip and substrate electrodes to investigate the conductance of single *Para* and *Meta* molecules in mesitylene (for details see Methods). The experiments generated thousands of conductance versus distance traces, some showing plateaux associated with the formation of single molecule junctions, followed by abrupt drops corresponding to the breakdown of the junctions. As plotted in Fig. 2a,b, the plateaux measured for *Para* are located at higher conductance values than those for *Meta*. For statistical analysis of the conductance, we constructed conductance histograms from about 700 such individual traces for the two molecules (Fig. 2c,d). The conductance histograms exhibit broad peaks, where the peak positions represent the average conductance values of single molecules—$(2.8 \pm 0.2) \times 10^{-4} G_0$ for *Para* and $(8.6 \pm 0.2) \times 10^{-5} G_0$ for *Meta*—where $G_0$ ($\frac{2e^2}{h} = 7.768 \times 10^{-5}$ S) is the conductance quantum.

Because both *Para* and *Meta* consist of three phenyl rings connected by C–C bonds, and terminated with iodine, the observed large conductance for *Para* and small conductance for *Meta* reflect constructive and destructive interference in the two molecules. As a control, we performed the same measurement in pure mesitylene in the absence of *Para* or *Meta*. In the control experiment, the conductance traces did not show obvious plateaux, and the conductance histogram shows no obvious peaks (Supplementary Fig. 2), confirming that the conductance peaks in Fig. 2c,d are from *Para* and *Meta*.

To further investigate the quantum interference effect on charge transport through the molecules, we determined the *I–V* characteristics of the two molecules. This was achieved by first bridging a molecule between the STM tip and the substrate electrodes, and then scanning the bias voltage between −2 V and 2 V over 0.1 s while recording the current through the molecule (see Methods)[29]. By repeating the procedure, we recorded over 10,000 such *I–V* curves and constructed a two-dimensional (2D) *I–V* histogram for each molecule from the intact *I–V* curves (see Supplementary Fig. 3 regarding the stability of the junctions at high bias voltages). The *I–V* curves show that the current increases linearly with the bias within a small bias voltage range and increases rapidly at high bias voltages for both *Para* and *Meta* (Fig. 3a,b). An important observation is that the linear range for *Para* is larger than that for *Meta*. This difference is more clearly observed in the 2D histograms of conductance versus bias voltage (Fig. 3c,d), which were obtained from the derivatives of the individual *I–V* curves. For *Para*, the 2D conductance histogram (Fig. 3c) shows a relatively shallow dependence of the conductance on the bias voltage. In contrast, the 2D conductance histogram of the *Meta* molecule exhibits a much steeper dependence (Fig. 3d). This observation reflects a key difference in the transmission functions of the *Para* and *Meta* molecules[30]. The average differential conductance within the linear range of the *I–V* curves agrees with the conductance measured at fixed low bias voltages (Supplementary Fig. 4). Nonlinearity occurs at high bias voltages, as discussed in Supplementary Fig. 5. Previous studies[14] have observed a V-shaped conductance versus bias curve for destructive interference, where the electrodes' Fermi level is close to anti-resonance. In this study, the *I–V* characteristics for *Meta* has a larger curvature near zero bias than that for *Para* but is not a pronounced V shape. This is due to the fact that the anti-resonance in the present system is far away from the electrode Fermi level (for more details see the calculation part). Tuning the Fermi level to anti-resonance is thus important for the clear observation of quantum interference.

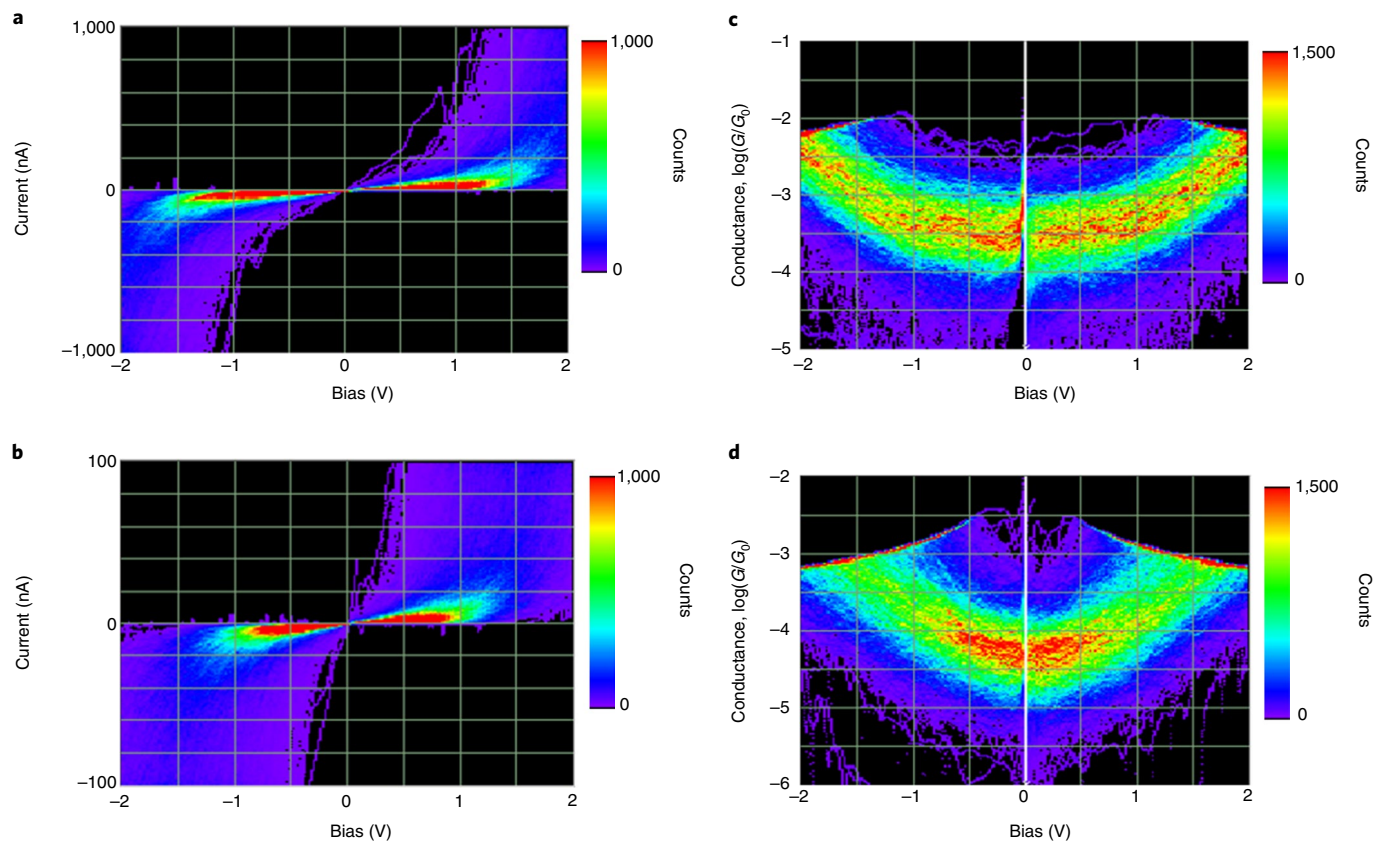

**Fig. 3 | I–V characteristics of *Para* and *Meta*. a,b**, 2D I–V histograms of *Para* (**a**) and *Meta* (**b**). **c,d**, 2D conductance–voltage (G–V) histograms of *Para* (**c**) and *Meta* (**d**). Each histogram consists of more than 1,000 curves.

To determine the transmission functions and examine the quantum interference versus energy, we shifted the energy alignment between the molecular orbital levels and the electrode Fermi level[31] with an electrochemical gate (Fig. 1a and Fig. 1d, top) (for experimental details see Methods). This changed the relative contributions of the HOMO and LUMO as described by the two terms in equation (2) and allowed us to obtain the transmission functions of the molecules (Fig. 1d, bottom). To apply an electrochemical gate voltage ($V_g$) to the molecules, we performed the measurement on *Para*/*Meta* adsorbed on the electrodes in 50 mM $KClO_4$ aqueous solution (the electrolyte is needed for electrochemical gating). The first step was to identify the formation of a single molecule junction from the conductance plateau, and then sweep the gate voltage from 0 to 0.3 V and from 0 to −0.5 V while recording the conductance of the molecule.

The *Para* conductance increases slowly from $2 \times 10^{-4} G_0$ to $4 \times 10^{-4} G_0$ with $V_g$ from −0.5 V to 0.3 V (Fig. 4a). The conductance range is consistent with that measured in mesitylene. The weak gate dependence indicates a relatively flat transmission function for *Para* within the energy range. In contrast, *Meta* exhibits a much stronger gate effect as the conductance increases from ~$6 \times 10^{-6} G_0$ at $V_g = -0.5$ V to ~$5 \times 10^{-4} G_0$ at $V_g = 0.3$ V (Fig. 4b). The strong gate effect reflects a sharp feature in the transmission function of *Meta*. The noise in the conductance during potential sweeps for both *Para* and *Meta* is due to thermal instability and mechanical drift during the sweep time of hundreds of milliseconds. Despite the noise, the observed overall difference between *Meta* and *Para* is consistent with the I–V characteristics of the two molecules. This experiment demonstrates that the quantum interference can be tuned by changing the relative coupling between the HOMO and LUMO with an external electrochemical gate. Applying an electrochemical gate voltage to a molecule is known to lead to oxidation or reduction of the molecules[3,12,15,31,32]. To show that the electrochemical gate control of quantum interference in the present system did not involve the redox reactions, we carried out cyclic voltammetry (CV) of both *Para* and *Meta* adsorbed on gold electrodes and did not observe the characteristic oxidation and reduction peaks associated with redox reactions (Supplementary Fig. 1f,g).

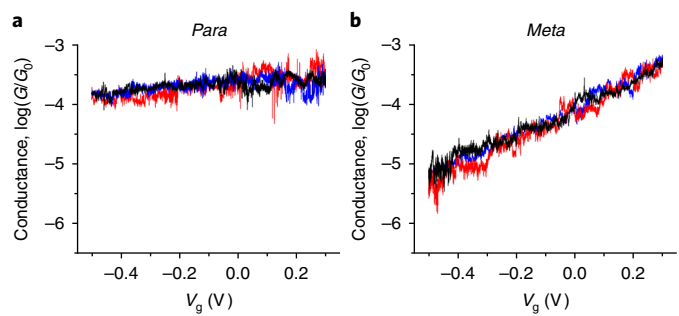

**Fig. 4 | Conductance of *Para* and *Meta* versus gate potential. a,b**, Conductance of *Para* (**a**) and *Meta* (**b**) versus gate potential ($V_g$) swept at 1 V s$^{-1}$, revealing a much larger gate dependency for the conductance of *Meta* than *Para*. Different colours represent different sweeps. The positive and negative traces are from separate sweeps.

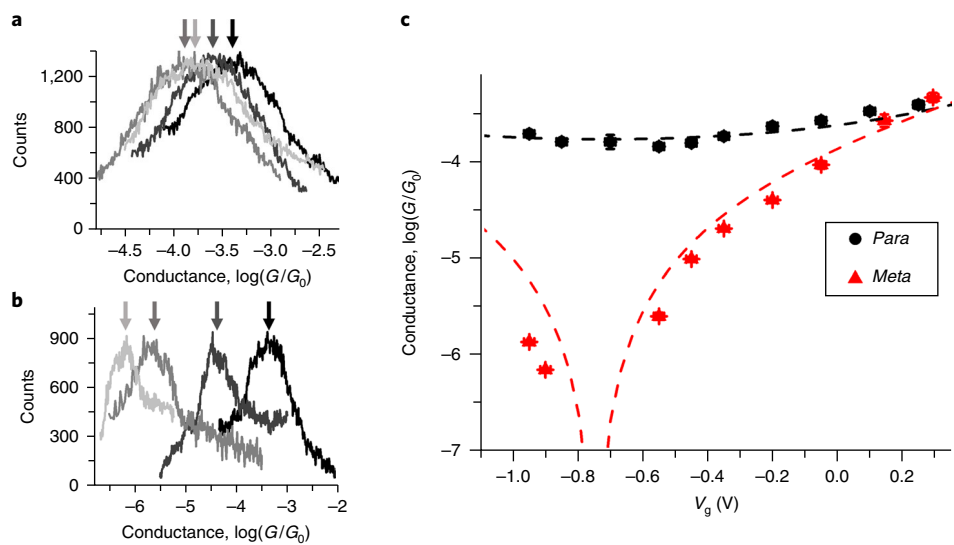

**Fig. 5 | Conductance of *Para* and *Meta* single molecules measured at different gate potentials. a**, Conductance histograms of *Para* measured at gate potentials of +0.25 V (black), −0.2 V (dark grey), −0.55 V (grey) and −0.95 V (light grey). **b**, Conductance histograms of *Meta* measured at gate potentials of +0.295 V (black), −0.2 V (dark grey), −0.55 V (grey) and −0.95 V (light grey). **c**, Conductance of *Para* (black circles) and *Meta* (red triangles). The *y* axis error bars represent standard deviations of three independent measurements; *x* axis error bars represent the potential shift of a quasi reference electrode. Black/red dashed lines are fits of the data by simple constructive and destructive interference transmission functions (for details see main text).

The single molecule conductance versus gate voltage measurements described so far required a stable molecular junction. We found that the molecular junctions became increasingly unstable at large gate voltages (negative or positive), which limited the range of the applied gate voltage. To statistically determine the conductance and transmission functions over a broader energy range, we performed small bias (0.1 V) conductance measurements quickly and repeatedly in the same manner as described in Fig. 2 and the Methods to obtain a large number of transient conductance curves at each fixed gate voltage and constructed conductance histograms at different gate voltages for both *Para* and *Meta*[33]. From the peaks in these conductance histograms we obtained the average conductance values of single molecules versus gate voltage. Because each measurement was transient, we were able to determine the conductance at gate voltage from 0.29 V to −0.95 V using this approach. For *Para*, the conductance histogram peak position varies mildly within a two- to three-fold change from $1.4 \times 10^{-4} G_0$ to $3.9 \times 10^{-4} G_0$ (Fig. 5a; complete 1D profiles, individual conductance decay curves and 2D conductance histograms are provided in Supplementary Figs. 6, 7 and 8). However, for *Meta*, when the gate voltage decreases from 0.295 V to −0.55 V, the conductance first decreases sharply from $4.7 \times 10^{-4} G_0$ to $2.5 \times 10^{-6} G_0$, and becomes too low to measure between −0.65 V and −0.85 V (Supplementary Fig. 9), then increases to the measurable range again from $6.8 \times 10^{-8} G_0$ to $1.4 \times 10^{-6} G_0$ at −0.9 V to −0.95 V (Fig. 5b; complete 1D profiles, individual conductance decay curves and 2D conductance histograms are presented in Supplementary Figs. 6, 7 and 10). Below −1 V, the junction formation yield is low due to instability of the junction (Supplementary Fig. 11). The large gate-induced conductance change (~200 times) and the clear conductance minimum near −0.75 V observed for *Meta* are distinctly different from the behaviours of *Para*. It is worth noting that the coupling between the molecules and the electrodes could change with the gate voltage, leading to a conductance change in *Meta*. However, the lack of gate dependence in the *Para* conductance shows that the change in the coupling is negligible.

Mechanical drift in a break junction could induce quantum interference[34,35]. To examine such mechanical effects, we determined the plateau slopes in conductance–distance traces at various gate voltages for both *Meta* and *Para* (Supplementary Fig. 12). The slope quantifies how the conductance changes on mechanical stretching, from which we observed little conductance change. Furthermore, the slope has the same sign on both sides of the conductance minimum, indicating that mechanical stretching is not a dominant factor in the observed large (~200 times) gate dependence of the conductance.

Figure 5c summarizes the measured conductance versus gate voltage for *Para* and *Meta*, from which we draw several important conclusions. First, the conductance increases with gate voltage from −0.55 V to 0.25 V for both *Para* and *Meta*, suggesting that the Fermi level is located closer to the HOMO level than to the LUMO level in the absence of gate voltage. This observation is supported by the theoretical prediction of energy alignment for phenyl substituted benzene compounds[11], and the energy alignment information allows us to compare the experimental data with the calculations. Note that a positive electrochemical gate potential shifts the molecular level up with respect to the Fermi level, which is opposite the behaviour in a traditional solid-state gate. Second, *Meta* exhibits a much stronger gate effect than *Para*, revealing a pronounced difference in the transmission function between destructive and constructive interference. The gating efficiency is described by subthreshold swing, which is the amount of gate voltage change needed to change the conductance by tenfold. In conventional semiconductor field-effect transistors, the subthreshold swing is limited by fundamental thermodynamics and is greater than ~60 mV at room temperature. In the present case, the subthreshold swing is determined by quantum interference, rather than thermodynamics, and the value estimated from the conductance change between −0.55 V and −0.45 V is 17 mV dec$^{-1}$, which is substantially smaller than in conventional devices[36–38]. Third, the sharp drop in the conductance of *Meta* as the gate voltage decreases to −0.55 V and the increase at more negative potentials after the conductance minimum, provide a direct observation of the signature anti-resonance in the transmission function. Finally, the conductance versus energy curves of *Para* and *Meta* can be fitted by the transmission functions determined with the simple model of equation (2) (dashed curves in Fig. 5c). We note that although electrochemical gating was used here to demonstrate the principle, other gating methods, such as electrostatic gating, could also be used to tune the quantum interference.

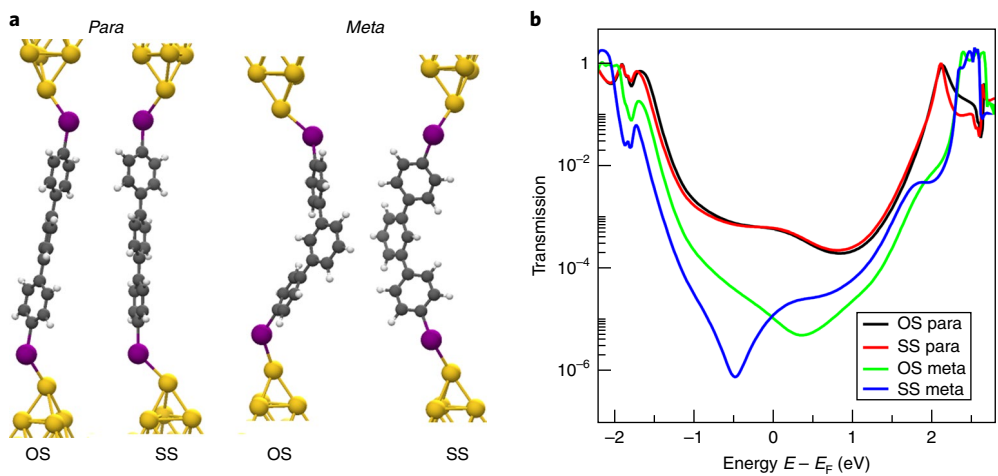

**Fig. 6 | Contact geometries and transmission functions. a**, Typical contact geometries considered for the transport calculations. OS represents the geometry with the dihedral angle between the two Au–I bonds at 0°, and SS represents the geometry with the dihedral angle at 180°. Au, yellow; I, purple; C, dark grey; H, white. **b**, Transmission function calculated for the different geometries of *Para* and *Meta* based on the DFT+Σ model.

To further validate the experimental data, we performed first-principles calculations of the *Para* and *Meta* molecules by combining the electronic structure derived from density functional theory (DFT) with non-equilibrium Green's functions (see Supplementary Methods and Supplementary Fig. 13) including an approximate self-energy correction for weakly coupled molecules (see Supplementary Methods and Supplementary Fig. 14) to account for over-estimation of the electronic conductance[39,40]. We show two typical molecule–electrode contact geometries for *Para* and *Meta* in Fig. 6a, where OS represents the geometry with the dihedral angle between the two Au–I bonds at 0°, and SS represents the geometry with that dihedral angle at 180°. The corresponding transmission functions of the two molecules with OS and SS geometries are given in Fig. 6b. The calculated HOMO–LUMO gaps are compared with UV–vis spectra in Supplementary Fig. 15. Because of the high efficiency of the electrochemical gating, we assume that the shift in transmission curves is roughly the same amplitude as the applied gate voltage[15,19,41]. In the absence of the gate voltage, the electrode Fermi energy was estimated from the DFT calculations using a slab model with bulk corrections[42], which was found to be −5.32 eV for a gold(111) surface. This value compares well with the experimentally determined workfunction of ~5.3 eV for the Au(111) surface[43,44]. At the Fermi energy, *Para* displays a more than one order of magnitude larger conductance ($G_{Para}^{DFT+\Sigma} = 5.8 \times 10^{-4} G_0$) than *Meta* ($G_{Meta}^{DFT+\Sigma} = 1.2 \times 10^{-5} G_0$), which agrees with the measured conductance for *Para*, but underestimates the one for *Meta*, although still within a reasonable range. The transmission functions of *Para* show a relatively small energy dependence within −1 and +1 V around $E_F$ for both OS and SS contact geometries. In contrast, the transmission functions of *Meta* show a much deeper energy dependence due to anti-resonance, which agrees with the experimental data. The transmission function of *Meta* is more sensitive to the contact geometry. For OS, the transmission spectrum is approximately symmetric with respect to the HOMO and LUMO resonance, and the anti-resonance is located in the centre of the HOMO–LUMO gap. However, the transmission spectrum for SS is asymmetric, with the anti-resonance shifted towards the HOMO. The exact contact geometries of the molecules remain unknown, but the first-principles calculations provide a good prediction of the basic anti-resonance feature regardless of the detailed geometry.

We have archived continuous control of quantum interference in single molecules with an external gate and studied its effect on the charge transport through the molecules. The gate voltage tunes the relative coupling strength between the HOMO and LUMO and leads to different gate dependences of the conductance for molecules with constructive and destructive interferences. By measuring the conductance versus gate voltage, we have determined the transmission functions of the molecules, providing direct observation of an anti-resonance in the transmission spectra, a key signature of quantum interference, and allowing a detailed comparison with theoretical calculations, including the shape and energy dependence of the transmission function. By controlling the quantum interference, we have further shown that conductance of a single molecule can be tuned over two orders of magnitude. Conventional semiconductor devices rely on gate control of carrier densities, which have a minimum subthreshold swing of ~60 mV dec$^{-1}$, a limit originating in thermodynamics. Our work shows that gate control of quantum interference leads to a subthreshold swing as low as ~17 mV dec$^{-1}$ at room temperature, much smaller than the values of conventional field-effect transistors. A low subthreshold swing is critical to high-speed and low-power electronics. The finding thus demonstrates a unique feature of quantum interference.

### Online content

Any methods, additional references, Nature Research reporting summaries, source data, statements of data availability and associated accession codes are available at https://doi.org/10.1038/s41563-018-0280-5.

### Acknowledgements
The authors (N.T. and Y.L.) thank D.N. Beratan and A. Nitzan for stimulating discussions. The authors acknowledge financial support from the National Natural Science Foundation of China (grants nos. 21773117 and 21575062, to H.W., Z.W.), the Ministry of Education, Culture, Sports, Science and Technology (MEXT), Japan (Grant-in-Aid for Scientific Research on Innovative Areas 'Molecular Architectonics: Orchestration of Single Molecules for Novel Functions'; grant no. 25110009, to Y.A and M.B.), the Japan Society for the Promotion of Science (Grant-in-Aid for Young Scientists (Start-up); KAKENHI grant no. 15H06889, to M.B.) and the National Natural Science Foundation of China (grants nos. 21674023 and 51722301, to G.L. and G.Z.).

### Author contributions
N.T., Y.L., L.X., G.Z., G.L., Y.A. and M.B. designed the research. G.L. and G.Z. synthesized the studied molecules. Y.L., A.R., H.W. and Z.W. performed and analysed the experiments. M.B., Y.A., D.R.B. and T.M. performed and analysed the DFT and transport calculations. Y.L., N.T., M.B. and G.L. wrote the paper. All authors contributed to revising the manuscript and agreed on its final content.

### Competing interests
The authors declare no competing interests.

### Additional information
**Supplementary information** is available for this paper at https://doi.org/10.1038/s41563-018-0280-5.

**Reprints and permissions information** is available at www.nature.com/reprints.

**Correspondence and requests for materials** should be addressed to Y.A., G.Z. or N.T.

**Publisher's note:** Springer Nature remains neutral with regard to jurisdictional claims in published maps and institutional affiliations.

## Methods

**Synthesis of *Para* and *Meta*.** All reactions and manipulations were carried out under a standard inert atmosphere and using Schlenk techniques. Anhydrous toluene was distilled from sodium benzophenone ketyl and anhydrous dichloromethane (DCM) was distilled from $CaH_2$. Stainless-steel syringes were used to transfer these moisture-sensitive solvents. 1,3-Dibromobenzene, 1,4-dibromobenzene, trimethylsilylphenylboronic acid and iodine monochloride were purchased from J&K Chemical. Proton nuclear magnetic resonance ($^1$H NMR, 400 MHz) spectra and carbon NMR ($^{13}$C NMR, 100 MHz) spectra were measured on a Varian Mercury Plus-400 spectrometer. The chemical shifts were in parts per million downfield from tetramethylsilane (TMS) and $CDCl_3$. The splitting patterns are designated as follows: s (singlet), d (doublet) and m (multiplet).

*Synthesis of 4,4″-bis(trimethylsilyl)-1,1′:3′,1″-terphenyl.* Under an argon atmosphere, a mixture of 1,3-dibromobenzene (1.00 g, 4.27 mmol), trimethylsilylphenylboronic acid (1.82 g, 9.40 mmol), $K_2CO_3$ (2.77 g, 20.44 mmol) and $Pd(PPh_3)_4$ (0.25 g, 0.22 mmol) in a mixed solvent of toluene (20 ml) and water (10 ml) was stirred and heated at 85 °C for 8 h. The mixture was cooled to room temperature, extracted with diethyl ether and washed with water. The separated organic layer was dried over anhydrous $MgSO_4$. After removal of the solvent by evaporation, the residue was purified by column chromatography on silica gel to afford the desired white product (0.76 g, yield 48%). $^1$H NMR (400 MHz, $CDCl_3$, δ ppm): 7.84 (s, 1H), 7.63 (s, 8H), 7.59 (d, $J = 7.2$ Hz, 2H), 7.51 (m, 1H), 0.32 (s, 18H). $^{13}$C NMR (100 MHz, $CDCl_3$, δ ppm): 141.96, 141.78, 139.65, 134.08, 129.40, 126.81, 126.42, 126.37, −0.85.

*Synthesis of 4,4″-diiodo-1,1′:3′,1″-terphenyl (Meta).* Iodine monochloride (2.5 ml of 1.0 M solution in $CH_2Cl_2$) was added to a solution of 4,4″-bis(trimethylsilyl)-1,1′:3′,1″-terphenyl (300 mg, 0.8 mmol) in $CH_2Cl_2$ (10 ml) at 0 °C under an argon atmosphere. The reaction mixture was stirred at room temperature for 12 h. $NaHSO_3$ aqueous solution was then added to remove unreacted iodine monochloride. The organic layer was separated and dried over anhydrous $MgSO_4$. The solvent was removed in a rotary vacuum evaporator, and the residue was chromatographed on a silica gel column to give white solid (220 mg, 58%). $^1$H NMR (400 MHz, $CDCl_3$, δ ppm): 7.78 (d, $J = 8.4$ Hz, 4H), 7.70 (s, 1H), 7.55–7.51 (m, 3H), 7.38 (d, $J = 8.4$ Hz, 4H). $^{13}$C NMR (100 MHz, $CDCl_3$, δ ppm): 141.05, 140.67, 138.15, 129.72, 129.28, 126.47, 125.77, 93.55.

*Synthesis of 4,4″-bis(trimethylsilyl)-1,1′:4′,1″-terphenyl.* Under an argon atmosphere, a mixture of 1,4-dibromobenzene (1.00 g, 4.27 mmol), trimethylsilylphenylboronic acid (1.82 g, 9.40 mmol), $K_2CO_3$ (2.77 g, 20.44 mmol) and $Pd(PPh_3)_4$ (0.25 g, 0.22 mmol) in a mixed solvent of toluene (20 ml) and water (10 ml) was stirred and heated at 85 °C for 8 h. The mixture was cooled to room temperature, extracted with diethyl ether and washed with water. The separated organic layer was dried over anhydrous $MgSO_4$. After evaporation of the solvent, the residue was purified by column chromatography on silica gel to afford the desired product (0.93 g, yield 58%). $^1$H NMR (400 MHz, $CDCl_3$, δ ppm): 7.70 (s, 4H), 7.65 (s, 8H), 0.33 (s, 18H). $^{13}$C NMR (100 MHz, $CDCl_3$, δ ppm): 141.30, 140.36, 139.59, 134.12, 127.75, 126.61, -0.82.

*Synthesis of 4,4″-diiodo-1,1′:4′,1″-terphenyl (Para).* Iodine monochloride (2.5 ml of 1.0 M solution in $CH_2Cl_2$) was added to a solution of 4,4″-bis(trimethylsilyl)-1,1′:3′,1″-terphenyl (300 mg, 0.8 mmol) in $CH_2Cl_2$ (10 ml) at 0 °C under an argon atmosphere. The reaction mixture was stirred at room temperature for 12 h. $NaHSO_3$ aqueous solution was then added to remove unreacted iodine monochloride. The organic layer was separated and dried over anhydrous $MgSO_4$. The solvent was removed in a rotary vacuum evaporator, and the residue was chromatographed on a silica gel column to give white solid (264 mg, 68%). $^1$H NMR (400 MHz, $CDCl_3$, δ ppm): 7.77 (d, $J = 8.4$ Hz, 4H), 7.63 (s, 4H), 7.38 (d, $J = 8.4$ Hz, 4H). The solubility of the target compound was too low to perform $^{13}$C NMR measurements. ($^1$H NMR spectra of *Meta* and *Para* in $CDCl_3$ are presented in Supplementary Figs. 16 and 17.)

**Immobilization of *Para* and *Meta* on gold electrodes.** *Para* and *Meta* were immobilized on a gold thin-film electrode (with ~160 nm gold on mica) prepared with a vapour deposition system under ultrahigh vacuum. Before each experiment, the gold substrate was briefly annealed with a hydrogen flame, immersed in mesitylene (Sigma-Aldrich, 98%) containing 30 μM *Para* or *Meta*, incubated overnight, then rinsed with mesitylene and dried with nitrogen gas. We bubbled the aqueous solution of 50 mM $KClO_4$ (Sigma-Aldrich, 99%) for 3 h to purge oxygen, then added the solution on top of the *Para*- or *Meta*-covered gold film in a glove box under a nitrogen atmosphere.

**Measurement of single molecule conductance.** Electrochemical gate-controlled break junction measurements were carried out with an electrochemical scanning tunnelling microscope (EC-STM) consisting of a controller (Nanoscope E, Digital Instruments), a STM scanner (Molecular Imaging) and a bipotentiostat (Agilent). For each experiment, the STM tip was freshly prepared by cutting a gold wire (0.25 mm diameter, 99.5%). For electrochemical gate control, the tip was coated with Apiezon wax to reduce ionic leakage current (<1 pA). A Ag wire and Pt coil were used as quasi-reference electrode and counter electrode, respectively. The potential drift of the Ag quasi reference is characterized in Supplementary Fig. 18.

We performed STM break junction measurements with the following three approaches. In the first, the sample was measured in a mesitylene solution of 30 μM *Para*/*Meta* without applying gate potential[18]. A small bias (0.1 V) was applied between the STM tip and the substrate electrodes. The STM tip was repeatedly brought into contact and retracted from the substrate, during which thousands of conductance–distance traces (Fig. 2a,b) were collected at a sampling rate of 10 kHz (Supplementary Fig. 19) to construct conductance histograms (Fig. 2c,d). In the second approach, a quasi-reference electrode (Ag wire) and a counter electrode (Pt coil) were inserted in 50 mM $KClO_4$ aqueous solution to allow gate-controlled conductance measurements. The break junction measurement was carried out in the same way as the first approach with the gate voltage held at a fixed value ranging from −1 V to 0.25 V (ref. [33]). In particular, the bias voltage applied between the STM tip and substrate was 0.2 V for the measurement of *Meta* at $V_g$ more negative than −0.55 V to enlarge the observable conductance window. Besides the measurement of *Meta* at $V_g < -0.55$ V, the bias applied in all other measurements was 0.1 V. The experiments were conducted in a glovebox under a nitrogen atmosphere. In the third approach, instead of applying a fixed gate voltage, we studied the conductance of a molecule while sweeping the gate voltage[31]. We fixed the tip–substrate bias at 0.1 V and the initial gate voltage at 0 V and performed break junction measurements. Once a plateau at the single molecule conductance level was detected automatically with a LabView program, signalling the formation of a single molecule junction, the tip was held in position and the potential was scanned at 1 V s$^{-1}$ to record conductance versus potential (Fig. 4).

**Measurement of single molecule I–V characteristics.** We first carried out the break junction measurements in mesitylene solution containing either *Para* or *Meta*. When a molecule formed the bridge between the STM tip and the gold film, a plateau appeared in the conductance versus distance trace, the tip was held in position and the bias was swept between −2 V to 2 V in 0.1 s to record a current–voltage characteristic curve at a sampling rate of 100 kHz (ref. [29]). We repeated the procedure to obtain thousands of individual current–voltage characteristic curves. From each current–voltage characteristic curve, we extracted a conductance–voltage curve. We constructed 2D histograms of current–voltage characteristics (Fig. 3a,b) and conductance–voltage curves (Fig. 3c,d) from more than 1,000 individual curves.

## Code availability
The DFT code CONQUEST is available at http://www.order-n.org and the corresponding module used to calculate the quantum transport properties is available from M.B. upon reasonable request.

## Data availability
The data that support the findings of this study are available from the corresponding author upon reasonable request.